\documentclass{ws-procs9x6}            
\newcommand{\bra}[1]{\langle#1\rvert}
\newcommand{\ket}[1]{\lvert#1\rangle}

\newcommand{\braopket}[3]{\langle #1 | #2 | #3\rangle}

\newcommand{\cA}{\mathcal{A}}
\newcommand{\cH}{\mathcal{H}}
\newcommand{\cO}{\mathcal{O}}
\DeclareMathOperator{\tr}{tr}
\DeclareMathOperator{\spa}{span}
\DeclareMathOperator{\sgn}{sgn}
\begin{document}
\title{Target Space Entanglement\\ in Quantum Mechanics of Fermions and Matrices}

\author{Sotaro Sugishita$^*$}

\address{Institute for Advanced Research, 
Nagoya University\\
and\\
Department of Physics, 
Nagoya University,\\
Nagoya, Aichi 464-8601, Japan\\
$^*$E-mail: sugishita.sotaro.r6@f.mail.nagoya-u.ac.jp}

\begin{abstract}
Quantum entanglement is closely related to the structure of spacetime in quantum gravity. 
For quantum field theories or statistical models, we usually consider base space entanglement. 
However, target space instead of base space sometimes directly connects to our spacetime. 
In these cases, it is natural to consider a concept of target space entanglement. 
To define the target space entanglement, we consider a generalized definition of entanglement entropy based on an algebraic approach. 
This approach is reviewed and is applied to the first quantized particles, in particular, fermions.
This article is based on the paper JHEP 08 (2021) 046\cite{Sugishita:2021vih}.
\end{abstract}


\bodymatter

\section{Introduction}
It is widely believed that 
quantum entanglement is closely related to the structure of spacetime in quantum gravity. 
In the AdS/CFT correspondence, the Ryu-Takayanagi formula\cite{Ryu:2006bv} states that entanglement about the base space in holographic CFTs is connected to the area of minimal surface in the bulk. 
As in this example,  we often consider the base space entanglement in quantum field theories or statistical models. 
However, target space instead of base space sometimes directly connects to our spacetime, for example, perturbative string theories or matrix models. 
Thus, it is natural to investigate a notion of target space entanglement  \cite{Mazenc:2019ety, Das:2020jhy, Das:2020xoa}. See also recent Refs.~\citenum{Hampapura:2020hfg, Sugishita:2021vih, Frenkel:2021yql, Tsuchiya:2022ffu, Das:2022mtb}.\footnote{A concept of entanglement in string theories (matrix models) is investigated in \citenum{Das:1995vj} and revisited in  \citenum{Hartnoll:2015fca}.}

In Ref.~\citenum{Mazenc:2019ety}, the target space entanglement is defined using an algebraic approach. 
We will review this approach in \sref{sec:alg}, and apply it to quantum mechanics of fermions in \sref{sec:QM}, \sref{sec:slater} and \sref{sec:circle}.

\section{Definition of entanglement entropy based on subalgebras of operators}
\label{sec:alg}
Let us recall the conventional definition of entanglement entropy (EE). 
Suppose that a total density matrix $\rho$ is given for a Hilbert space $\mathcal{H}=\mathcal{H}_B \otimes \mathcal{H}_{\bar{B}}$.
The EE for subsystem $\mathcal{H}_B$ is defined as the von Neumann entropy of the reduced density matrix $\rho_B=\tr_{\bar{B}}\rho$ as $S_B=-\tr_{B} \rho_B \log \rho_B$.
This definition relies on the tensor product structure of the Hilbert space, $\mathcal{H}=\mathcal{H}_B \otimes \mathcal{H}_{\bar{B}}$.
However, total Hilbert spaces sometimes do not have such simple tensor-factorized forms.
For example, the Hilbert space of a first-quantized (non-relativistic) particle in a  space $\mathbb{R}^d$ is schematically given by a ``direct sum'' as $\mathcal{H}=\spa\{\ket{x}|\,x\in \mathbb{R}^d\}$.
Thus, even if we divide the space $\mathbb{R}^d$ into two subregions $\mathbb{R}^d=B \cup \bar{B}$, it is difficult to take the ``partial trace'' on $\bar{B}$.

The algebraic approach enables us to define EE without relying on the tensor product structure  (see, e.g.,  the references\cite{ohya2004quantum, Casini:2013rba, Harlow:2016vwg, Sugishita:2021vih}). 
The algebraic definition is based on the subalgebra of operators (observables).
If a total density matrix $\rho$ is given, and we have a restricted set of operators (subalgebra $\cA$), an entropy $S_{\cA}(\rho)$ associated with the subalgebra $\cA$ is defined.
This concept is natural, if we recall the meaning of entropy in information theory.
The entropy is a measure of uncertainty about the whole information when we can only know partial information. 
If an observer can use only a subset of operators $\cA$, the whole information is not obtained. 
Entropy $S_{\cA}(\rho)$ quantifies the amount of uncertainty (or unknownness).
In this sense, the usual EE, $S_B=-\tr_{B} \rho_B \log \rho_B$, for $\mathcal{H}=\mathcal{H}_B \otimes \mathcal{H}_{\bar{B}}$ represents uncertainty for an observer who can probe only subsystem $\mathcal{H}_B$. That is, it is the entropy for the subalgebra $\mathcal{L}(\mathcal{H}_B) \otimes 1_{ \mathcal{H}_{\bar{B}}}$.\footnote{Here, $\mathcal{L}(V)$ denotes a set of linear operators on linear space $V$, and $1_V$ does the identity operator on $V$.}
The choice of subalgebra $\cA$ is arbitrary, and we do not need the tensor product structure.

For general subalgebra $\cA$, 
the entropy $S_{\cA}(\rho)$ is computed as follows. 
First, the `reduced density matrix' $\rho_{\cA}$ is uniquely determined from $\rho$ and $\cA$ as an operator in $\cA$ satisfying the following equation:
\begin{align}
    \tr(\rho_{\cA} \cO)=\tr(\rho_{\cA} \rho), \qquad ~^{\forall}\cO \in \cA.
    \label{def:rhoA}
\end{align}
For example, 
if the total Hilbert space has a tensor product form as $\mathcal{H}=\mathcal{H}_B \otimes \mathcal{H}_{\bar{B}}$, and we take the subalgebra $\cA$ as $\cA=\mathcal{L}(\mathcal{H}_B) \otimes 1_{ \mathcal{H}_{\bar{B}}}$, then $\rho_{\cA}$ is given by $\rho_B \otimes 1_{ \mathcal{H}_{\bar{B}}}/\dim \mathcal{H}_{\bar{B}}$.
The point is that the definition \eref{def:rhoA} is applicable even when the Hilbert space does  not have the tensor product structure.

Furthermore, for a given subalgebra, we can decompose the Hilbert space into blocks of tensor products where the subalgebra acts nontrivially only on each tensor component as follows:
\begin{align}
\label{decomp}
    \cH=\bigoplus_k \cH_{B_k}\otimes \cH_{\bar{B}_k} \quad \mathrm{s.t.}
    \quad\cA=\bigoplus_k  \mathcal{L}(\cH_{B_k})\otimes 1_{\bar{B}_k}.
\end{align}
This decomposition is uniquely fixed by the subalgebra $\cA$.
We represents the projection onto each block by $\Pi_k$.
We define the density matrix $\rho_k$ on the projected space $\Pi_k \cH$ as
\begin{align}
\label{rho_k}
    \rho_k:=\frac{1}{p_k}\Pi_k \rho \Pi_k,
\end{align}
where $p_k$ is a normalization factor defined as $p_k:=\tr(\Pi_k \rho \Pi_k)$ and is a probability of being in the sector  $\Pi_k \cH$ for the given $\rho$.
Since the projected space $\Pi_k \cH$ has a simple tensor-factorized form as $\Pi_k \cH=\cH_{B_k}\otimes \cH_{\bar{B}_k}$ in the decomposition \eqref{decomp}, we can consider the reduced density matrix of $\rho_k$ on $\cH_{B_k}$ as 
\begin{align}
    \rho_{B_k}:= \tr_{\bar{B}_k}\rho_k.
\end{align}
Then,  the `reduced density matrix' $\rho_{\cA}$ satisfying \eref{def:rhoA} is given by
\begin{align}
    \rho_{\cA}=\bigoplus_k p_k\, \rho_{B_k} \otimes \frac{ 1_{\bar{B}_k}}{\dim(\cH_{\bar{B}_k})}.
\end{align}

We define the reduced density matrix $\rho_{B}$ on space $\cH_B=\bigoplus_k \cH_{B_k}$ as 
\begin{align}
\label{rhoB}
    \rho_{B}:=\bigoplus_k p_k \rho_{B_k}.
\end{align}
EE $S_{\cA}(\rho)$ is defined as the von Neumann entropy
\begin{align}
    S_{\cA}(\rho)=-\tr_B \rho_B \log \rho_B=-\sum_k p_k \log p_k+\sum_k p_k S(\rho_{B_k}),
    \label{EE}
\end{align}
where $S(\rho_{B_k}):=-\tr_{B_k}\rho_{B_k}\log \rho_{B_k}$.
The first term in  the r.h.s. of \eref{EE} is called the classical part,
\begin{align}
    S_{cl}(\rho,\cA):=-\sum_k p_k \log p_k,
\end{align}
and is the Shannon entropy of the probability distribution $\{p_k\}$.
On the other hand, the second term in  the r.h.s. of \eref{EE}
is called the quantum part $S_{q}(\rho,\cA)$.
The expression in \eref{EE} is similar to  the symmetry resolved entanglement entropy  \cite{Goldstein:2017bua, Bonsignori:2019naz}.

\subsection{Example: Entanglement in a single qubit}
As a concrete example of EE in the algebraic approach, 
we consider a single qubit.
The Hilbert space is two-dimensional space, $\cH=\spa\{\ket{0}, \ket{1}\}$.
We usually  consider entanglement between two qubits. 
The algebraic approach enables us to consider ``EE'' even for a single qubit.

The full set of operators $\mathcal{L}(\cH)$ is $\mathcal{L}(\cH)=\spa\{I,\sigma_x,\sigma_y,\sigma_z\}$.\footnote{We take a basis such that $\sigma_z=\begin{pmatrix}
1&0\\
0&-1
\end{pmatrix}$ with $\sigma_z\ket{k}=(-1)^k\ket{k}$ $(k=0,1)$.}
If we take the subalgebra as this full algebra, the decomposition \eref{decomp} is trivial as
\begin{align}
    \cH=\cH\otimes \mathbb{C}
\end{align}
with $\cA=\mathcal{L}(\cH)\otimes 1$.
In this case, $\rho_B$ in \eref{rhoB} is just the original $\rho$. 
Thus, the EE associated with the full algebra $\mathcal{L}(\cH)$ is just the von Neumann entropy of $\rho$, 
\begin{align}
    S_{\mathcal{L}(\cH)}(\rho)=-\tr \rho \log \rho.
\end{align}
In particular, if state $\rho$ is pure, the entropy vanishes as $ S_{\mathcal{L}(\cH)}(\rho)=0$.
It means that the pure state is not ambiguous and is completely determined by quantum tomography if we can use any operators. 

Situation changes when we can use only a subset of operators. 
Let us suppose that we can probe only $z$-direction. 
This corresponds to taking subalgebra $\cA=\spa\{1,\sigma_z\}$.
The decomposition \eqref{decomp} for this choice of the subalgebra is
$\cH=\spa\{\ket{0}\} \oplus \spa\{\ket{1}\}$ where $\cA=\spa\{1,\sigma_z\}$ can be represented as $\cA=\spa\left\{\begin{pmatrix}
1&0\\
0&0
\end{pmatrix}
\right\}
\oplus
\spa\left\{\begin{pmatrix}
0&0\\
0&1
\end{pmatrix}
\right\}
$.
The projection $\Pi_k$ are  $\Pi_k=\ket{k}\bra{k}$ $(k=0,1)$.
We then have $p_k=\braopket{k}{\rho}{k}$ and  $\rho_{B_k}=\Pi_k$.
Thus, the EE associated with the subalgebra $\cA$ is
\begin{align}
    S_{\cA}(\rho)=-p_0 \log p_0-p_1 \log p_1,
\end{align}
where the quantum part $S_q(\rho, \cA)$ always vanishes, and the entropy is just the classical Shannon entropy of the probability distribution that the qubit is measured in $0$ or $1$ for the given state $\rho$.
Even pure states in general have non-vanishing entropy (except for the case where states are eigenstates of $\sigma_z$).
The non-vanishing entropy reflects the fact that pure states are ambiguous for restricted observers who can probe only $z$-direction. 
In fact, the observers cannot distinguish pure states with mixed states $\rho=\begin{pmatrix}
p_0&0\\
0&p_1
\end{pmatrix}$.

\section{Entanglement of fermions with a fixed number}\label{sec:QM}
We now consider target space entanglement of first-quantized $N$ fermions by the algebraic approach.
The Hilbert space of the single particle is represented by $\cH^{(1)}$.  
It is given by $\cH^{(1)}=\spa\{\ket{x}|x\in M\}$ where $M$ is the target space of particles.
The Hilbert space $\cH^{(N)}$ of $N$ fermions is given by the $N$-th exterior power of $\cH^{(1)}$ as 
\begin{align}
    \cH^{(N)}~=\bigwedge\nolimits^{N}\cH^{(1)}, 
\end{align}
which is spanned as $\cH^{(N)}=\spa\{\ket{x_1}\wedge\dots \wedge\ket{x_N}|x_i\in M (i=1,\dots,N)\}$.

We take a subregion $B$ in the target space $M$, and consider the EE of this subregion. 
Since the Hilbert space $\cH^{(N)}$ does not have a tensor-factorized structure with respect to the target space coordinates, we adopt the algebraic approach instead of the conventional definition.
The subalgebra we take is the set of operators acting non-trivially only on particles in subregion $B$, which is represented by $\cA(B)$. For example, when $N=1$, the subalgebra $\cA(B)$ is given by $\cA(B)=\spa\left\{\ket{y}\bra{y'}|y,y'\in B \right\}\oplus\spa\left\{\int_{\bar{B}}dz \ket{z}\bra{z} \right\}$ where $\bar{B}$ is the complement region of $B$. 
For general $N$, we can decompose $\cH^{(N)}$ into a direct sum of the following subsectors as 
\begin{align}
\label{N=k+N-k}
    \cH^{(N)}=\bigoplus_{k=0}^N \cH^{(N)}_k.
\end{align}
The subsector $\cH^{(N)}_k$ consists of states where $k$ particles in $B$ and $N-k$ ones in $\bar{B}$ as 
\begin{align}
    \cH^{(N)}_k=\spa\left\{\ket{x_1}\wedge\dots\wedge \ket{x_{N}}\,
    |\,x_1,\dots, x_k \in B, \, x_{k+1},\dots, x_{N} \in \bar{B}
    \right\}.
\end{align}
To represent the subalgebra $\cA(B)$, we introduce the following abbreviated notation:
\begin{align}
   & \ket{\{x\}_n}=\ket{x_1}\wedge \dots \wedge \ket{x_n} \in \bigwedge\nolimits^{n}\cH^{(1)}. 
\end{align}
The subalgebra $\cA(B)$ is then given by 
\begin{align}
    \cA(B)=\bigoplus_{k=0}^N \cA_k,
\end{align}
where $\cA_k$ is a subalgebra on $ \cH^{(N)}_k$ and takes the form
\begin{align}
    \cA_k=&\spa\left\{\int_{\bar{B}} dz_1\dots dz_{N-k}
    (\ket{\{y\}_k}\wedge \ket{\{z\}_{N-k}}) (\bra{\{y'\}_k}\wedge \bra{\{z\}_{N-k}})  \right\}
    \nonumber
    \\
    &\text{with} \quad y_1, \dots, y_k, y'_1,\dots, y'_k \in B.
\end{align}

Since the subalgebra $\cA(B)$ is specified, we can compute the entropy $S_{\cA(B)}$ associated with this subalgebra in the manner described in the previous section.\footnote{In this case, the projection $\Pi_k$ in \eref{rho_k} is the projection to the subsector $\cH^{(N)}_k$ in \eref{N=k+N-k}.} 
We call this entropy the target space entanglement entropy $S_B$ because the subalgebra $\cA(B)$ is characterized by the subregion $B$ in the target space of particles.
In the second quantized picture, we can define the conventional entanglement entropy for subregion $B$. 
The target space EE $S_B$ agrees with this base space EE\cite{Mazenc:2019ety, Das:2020jhy, Sugishita:2021vih}.

\section{Fermions in the Slater determinant states}\label{sec:slater}
To be more specific, we focus on pure states whose $N$-body wave functions are given by the Slater determinants as 
\begin{align}
    \psi(x_1,\dots,x_N)=\frac{1}{\sqrt{N!}}\det[\chi_i(x_j)] \qquad (i,j=1,\dots ,N),
\end{align}
where $\chi_i(x)$ are  one-body wave functions normalized as
\begin{align}
   \int_M dx\,  \chi^\ast_i(x) \chi_j(x)=\delta_{ij}.
\end{align}

The target space EE for subregion $B$ can be evaluated as \eref{EE} by computing $p_k$ and $S(\rho_{B_k})$ for the pure states $\psi$.
After some computations (see Ref.~\citenum{Sugishita:2021vih} for details), we can find that the entropy $S_B$ follows the simple formula:
\begin{align}
\label{formula}
    S_B(\psi)=-\tr [X\log X+(1_N-X)\log (1_N-X)],
\end{align}
where $X$ is a $N\times N$ matrix given by
\begin{align}
    X_{ij}=\int_B dx\,  \chi^\ast_i(x) \chi_j(x).
\end{align}
We call $X$ overlap matrix. 
It is easy to show that the eigenvalues $\lambda_i$ of the overlap matrix  are in the range $0\leq \lambda_i \leq 1$.

From the formula \eqref{formula}, we can find that the entropy has the upper bound\footnote{We can also confirm that the classical part $S_{cl}$ is bounded as $S_{cl}(\rho;A)\lesssim \mathcal{O}(\log N)$.\cite{Sugishita:2021vih}} as 
\begin{align}
    S_B(\psi)\leq N\log 2.
\end{align}
The maximum entropy $N\log 2$ is proportional to the number of particles $N$, and thus follows an extensive property like thermal entropy. 
However, this upper bound is too generic. 
We expect that EE for ground states is not extensive but sub-extensive in local models.
In fact, we will see in the next section that 
the entropy of a ground state of $N$ free fermions behaves as $S\sim \mathcal{O}(\log N)$ in the large $N$ limit.

\section{Entanglement for free fermions in a circle}\label{sec:circle}
We now apply the formula \eqref{formula} to $N$ free fermions in a circle with length $L$, \textit{i.e.}, the target space $M$ is a circle.
The Hamiltonian is given by $H=\sum_{i=1}^N \frac{p_i^2}{2m}$, and we consider its ground state.
The one-body eigenfunctions are given by
$\chi_n(x)=\frac{1}{\sqrt{L}}e^{\frac{2\pi i}{L}n x}$ where $n$ are integers.
Supposing that the total number of particles $N$ is odd ($N=2K+1$), the $N$-body wave function for the ground state is given by the Slater determinant as
\begin{align}
    \psi(x_1,\cdots, x_{N})=\frac{1}{\sqrt{N!}}\sum_{\sigma \in S_N}\sgn(\sigma) \chi_{-K}(x_{\sigma(1)})\cdots\chi_{K}(x_{\sigma(N)}).
\end{align}
Thus, the target space entanglement for a subregion $B$ can be obtained by the formula \eqref{formula} with the $N\times N$ overlap matrix 
\begin{align}
      X_{n n'}=\int_B dx\,  \chi^\ast_n(x) \chi_{n'}(x) ,
\end{align}
where $n, n'$ runs in $-K, \dots, K$.

\subsection{Single interval}
In this subsection, we consider the case where the subregion $B$ is a single interval $I_1$ in the circle. 
We parameterize the length of the interval as $r L$ $(0\leq r \leq 1)$, \textit{i.e.}, $r$ is the ratio of the interval to the circle.

\begin{figure}
\centering
\includegraphics[width=3in]{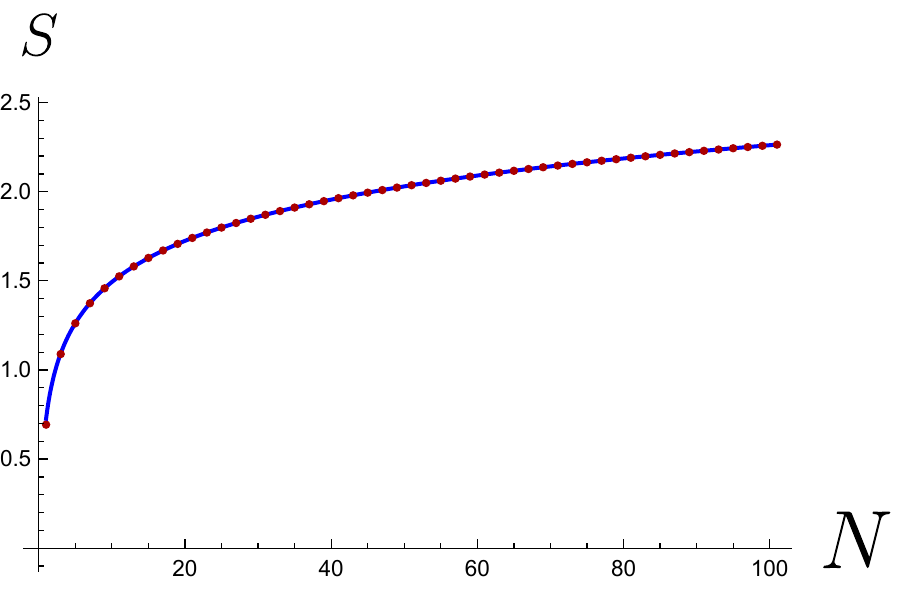}
\caption{EE for the half region. The red dots are the EE $S$ for $N=1,3, \cdots, 101$. The blue curve represents the large $N$ result \eqref{EE:sing-large} with $r=1/2$.}
\label{fig:half}
\end{figure}

In the large $N$ limit, the asymptotic behavior of the entropy can be obtained as
\begin{align}
\label{EE:sing-large}
S_{I_1} \sim \frac{1}{3} \log  [2N \sin(\pi 
r)]
+\Upsilon_1 
\end{align}
with 
\begin{align}
    \Upsilon_1=i\int^{\infty}_{-\infty}dw \frac{\pi w}{\cosh^2(\pi w)}\log \frac{\Gamma\left(\frac{1}{2}+i w\right)}{\Gamma\left(\frac{1}{2}-i w\right)}\sim 0.495018.
\end{align}
We show the plot of the entropy with the large $N$ result \eqref{EE:sing-large} in 
\fref{fig:half}.
It shows that the entropy is sub-extensive (not proportional to $N$).
Furthermore, the large $N$ behavior \eref{EE:sing-large} agrees with the EE for the single interval in $c=1$ CFTs on the circle\cite{Calabrese:2004eu} if we regard $N$ as a (dimensionless) cutoff.

\subsection{Entanglement entropy and mutual information for two intervals}
In this subsection, we consider two disjoint intervals $I_1$ and $I_2$ in the circle. 
Suppose that the coordinates of the circle is $x$ moving in $-\frac{L}{2} \leq x\leq \frac{L}{2}$ with the periodic condition $x\sim x+L$.
We take the two intervals as $I_1=\left(-\frac{d+r}{2}L,-\frac{d-r}{2}L\right)$ and $I_2=\left(\frac{d-r}{2}L,\frac{d+r}{2}L\right)$.

The EE for the subregion $I_1 \cup I_2$ can be analytically computed in the large $N$ limit\footnote{See Ref. \citenum{Sugishita:2021vih}.} as 
\begin{align}
\label{twointEE}
        S_{I_1\cup I_2}\sim  \frac{1}{3}\left[2\log[2N\sin (\pi r)]+\log\frac{ \sin[\pi (d+r)] \sin[\pi (d-r)]}{\sin^2(\pi d)}\right] +2\Upsilon_1.
\end{align}

We can also evaluated the target space mutual information; 
\begin{align}
\label{mi}
  I(I_1;I_2):=  S(I_1)+S(I_2)-S(I_1\cup I_2).
\end{align}
The large $N$ behavior is 
\begin{align}
\label{mi_largeN}
    I(I_1;I_2)\sim \frac{1}{3}\log\frac{\sin^2(\pi d)}
    { \sin[\pi (d+r)] \sin[\pi (d-r)]}.
\end{align}
The mutual information is finite even in the large $N$ limit. 
In addition, \eref{mi_largeN} agrees with the result in a $c=1$ CFT (free compact boson at the self-dual radius\cite{Calabrese:2009ez}), although the reason is not understood well.

\begin{figure}
\centering
\includegraphics[width=3in]{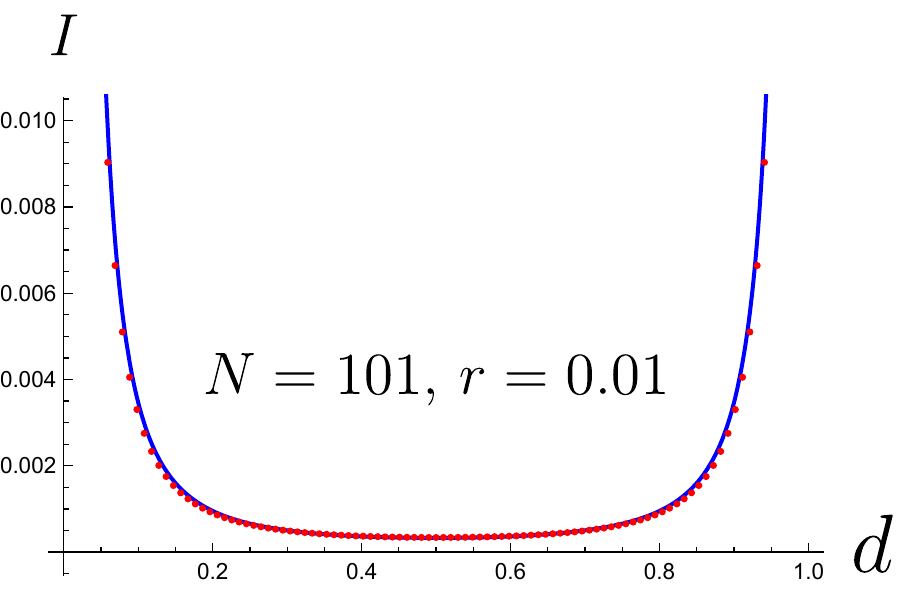}
\caption{Mutual information for two intervals. We take $N=101$ and set the parameter $r$ as $r=0.01$ (length of each interval is $rL$). The red dots represent  the mutual information for some values of $d$. The blue curve represents the large $N$ result \eqref{mi_largeN}.}
\label{fig:mi}
\end{figure}

The plot of the target space mutual information \eqref{mi} is \fref{fig:mi}.

\section{Brief conclusion}
The algebraic approach is a powerful method of characterizing entanglement. 
This approach might be useful beyond the target space entanglement. 
A similar idea to define entropy based on observables is also investigated as the observational entropy (see, \textit{e.g.}, Ref.~\citenum{Safranek:2020tgg}).

We have used the algebraic approach to define the target space entanglement of particles. 
In particular, we consider non-interacting fermions, which can be regarded as the singlet sectors of one-matrix models.
It is more interesting to consider entanglement in multi-matrix models, and its relation to holography.

\section*{Acknowledgments}
SS thanks the organizers of East Asia Joint Symposium on Fields and Strings 2021 for the opportunity to present the talk at Osaka City University. 
SS  acknowledges  support from JSPS KAKENHI Grant Number JP 21K13927.

\bibliographystyle{ws-procs9x6} 
\bibliography{ref}

\end{document}